\documentclass[12pt,prb,aps]{revtex4-1}
\usepackage{amssymb,amsmath}
\usepackage{graphicx}
\allowdisplaybreaks[1]

\begin{document}

\title{Two-Fluid Nonlinear Theory of Response of Tokamak Plasma to Resonant Magnetic Perturbation
\\[0.5ex]
~\\[0.5ex]
{\rm Richard Fitzpatrick}\\[0.5ex]
{\it Department of Physics}\\
{\it Institute for Fusion Studies}\\
{\it University of Texas at Austin}\\
{\it Austin TX, 78712, USA}~\\~\\}
\begin{abstract}
A comprehensive  two-fluid nonlinear theory of magnetic reconnection driven at a single, tearing-stable, rational
surface embedded in an H-mode tokamak plasma is presented.  The surface is assumed to be resonant with one of the dominant helical harmonics of an applied resonant
magnetic perturbation (RMP). The theory described in this paper is highly relevant to the problem of understanding the physics of
RMP-induced edge localized mode (ELM) suppression in tokamak plasmas. 
\end{abstract}
\maketitle

\section{Introduction}\label{s1}
Tokamak discharges operating in high-confinement mode (H-mode)\cite{wagner} exhibit intermittent bursts of heat and particle transport, 
emanating from the outer regions of the plasma, that are known as {\em edge localized modes} (ELMs).\cite{zohm} 
ELMs are fairly benign in present-day tokamaks. However, it is estimated that the heat load that ELMs
will deliver to the plasma-facing components in the divertor of a reactor-scale tokamak will be large enough to
unacceptably limit the lifetimes of these components via erosion.\cite{loarte}

The most promising method for the control of ELMs is via the application of static  {\em resonant magnetic perturbations}\/  (RMPs). Complete RMP-induced ELM suppression was first demonstrated in the DIII-D tokamak via the application of static, non-axisymmetric, 
magnetic fields with toroidal mode number $n_\varphi=3$.\cite{evans} Subsequently, either mitigation or compete suppression of
ELMs has been demonstrated on the JET,\cite{jet} ASDEX-U,\cite{asdex} KSTAR,\cite{kstar} and EAST\,\cite{east} tokamaks. Furthermore, the parameter range over which ELM suppression has been
observed in DIII-D has been extended to include low-collisionality ITER-shaped plasmas,\cite{lowc} hybrid scenarios, \cite{hybrid} and plasmas
with low, ITER-relevant, neutral beam momentum injection.\cite{lowt} These achievements have led to increased confidence that the
RMP ELM control technique can operate effectively in ITER.\cite{iter} 

At present, the physical mechanism of RMP-induced ELM suppression is not fully understood. ELMs are generally thought to
be caused by peeling-ballooning instabilities with intermediate toroidal mode numbers ($3<n_\varphi<20$) that are
driven by the strong pressure gradients and current densities characteristic of the edge region of an H-mode discharge, which is known as the {\em pedestal}.\cite{conner}
Consequently, early attempts to understand RMP-induced ELM suppression focused on the role of RMPs in reducing the
pressure gradient, and, thereby, reducing the bootstrap current density, in the pedestal. In particular,
the initial observations of ELM suppression were interpreted as an indication that the magnetic field in the pedestal 
had been rendered stochastic by the applied RMP, leading to greatly enhanced transport via thermal diffusion along
magnetic field-lines.\cite{evans,fenstermacher} This simplistic explanation was quickly abandoned because no
reduction in the electron temperature gradient in the pedestal is observed during RMP ELM suppression experiments,
whereas a very significant reduction would be expected in the presence of stochastic fields. It is now generally accepted that
response currents generated within the plasma play a crucial role in the perturbed equilibrium in the presence of RMPs, and that these currents  act to prevent the formation of magnetic islands---a process known as {\em shielding}---and, thereby, significantly reduce the stochasticity of
the magnetic field.\cite{berc} 

Current thinking suggests that density and temperature gradients in the pedestal are fixed by
stiff transport due to high-$n_\varphi$ instabilities, and that the  pedestal  grows in radial extent,
characterized by these fixed gradients, until the peeling/ballooning stability threshold is reached, and an
ELM is triggered.\cite{synder} According to this scenario, ELM suppression is achieved by limiting the expansion
of the pedestal before the peeling/ballooning stability threshold is crossed.\cite{snyder} It is hypothesized that pedestal
expansion is halted by a region of enhanced transport at the top of the pedestal that is not suppressed by ${\bf E}\times {\bf B}$ velocity shear, and
that this region is associated with the presence of an RMP-driven magnetic island chain located just outside, but  close to the top of, the pedestal.\cite{wade}

The aim of this paper is to develop a physical understanding of the dynamics of magnetic reconnection
driven at a single rational surface that is resonant with one of the dominant helical harmonics of an applied RMP. 
 The rational surface is assumed to be located just outside, but close to the top of, the pedestal of an H-mode tokamak plasma. The rational surface is also assumed to be intrinsically tearing-stable, so that
 any magnetic reconnection that occurs at the surface is due to the action of the applied RMP. Dealing with a single rational surface is a reasonable approach because the aforementioned lack of magnetic
 stochasticity in the pedestal indicates that any magnetic island chains  driven in this region are
 sufficiently narrow that they do not overlap with one another, which suggests that they evolve independently, at least to a first approximation. We are also concentrating on a particular rational
 surface (i.e., one located close to the top of the pedestal, and resonant with a dominant RMP harmonic) because there is ample evidence from experimental
 $q_{95}$ scans that RMP-induced ELM suppression depends crucially on the existence of this surface.\cite{lowc,asdex,wade} 
 
 This paper does not  explicitly address  the mechanism by which driven reconnection at the aforementioned
 rational surface leads to ELM suppression. However, there are two obvious candidate mechanisms. 
 The first is the degradation of radial confinement due to the fact that heat and particles can flow around the magnetic separatrix from one side of the island chain to the other.\cite{chang} The second is the modification of 
 the local plasma flow induced by the electromagnetic locking torque exerted on the island
 chain by the RMP. (A change in the flow affects the local ${\bf E}\times {\bf B}$ velocity shear which, in turn, can modify the local transport.) However, it is clear that both of these mechanisms are only operative if a relatively
 wide island chain is driven at the rational surface. In other words, they  only
work effectively if strong shielding breaks down at
 the rational surface. Hence, this paper will investigate the circumstances  in which the breakdown of shielding occurs.  
 
 Over the years, many different theoretical approaches have been taken to investigate driven reconnection at
 an intrinsically stable rational surface embedded in a tokamak plasma. These include single-fluid linear response models,\cite{rf1,rf2,curv} two-fluid linear response models,\cite{wael1,colef,wael} kinetic linear response models,\cite{heyn} single-fluid nonlinear response models,\cite{rf2} and two-fluid nonlinear response models.\cite{rf3} 
 
 Generally speaking, the aforementioned response models
 predict the same phenomenology.  (Here it is assumed that the linear models are used to calculate a quasi-linear electromagnetic locking torque that is balanced against a viscous restoring torque.) There exists a {\em shielded state}\/ in which driven magnetic
 reconnection at the rational surface is strongly suppressed by plasma flow, and a {\em penetrated state}\/
 in which the plasma flow is modified in such a manner as to allow significant magnetic reconnection. 
 As the resonant component of the RMP is gradually ramped up, a bifurcation from the 
 shielded to the penetrated state is triggered once the resonant component exceeds a critical {\em penetration threshold}. If the resonant component is then gradually ramped down then a bifurcation from the penetrated
 to the shielded state is triggered once the resonant component falls below a critical {\em de-penetration threshold}. However, the de-penetration threshold is usually significantly smaller than the penetration threshold. 
 Both bifurcations are accompanied by sudden changes in the plasma flow at the rational surface, which are, in turn, 
induced by changes in the electromagnetic locking torque exerted in the vicinity of the surface by the resonant component of the RMP. 

Despite the discussion in the previous paragraph, it is important to appreciate that there exist significant differences between the predictions of the
various response models. Single-fluid linear and single-fluid nonlinear models predict that the penetrated state is achieved when the ${\bf E}\times
{\bf B}$ velocity is reduced to zero at the rational surface.\cite{rf1,rf2} On the other hand, two-fluid linear theories predict that the
penetrated state is achieved when the perpendicular electron fluid velocity  is reduced to zero at the rational surface.\cite{colef}
Finally, two-fluid nonlinear theories predict that the
penetrated state is achieved when a velocity that is offset from the  ${\bf E}\times
{\bf B}$ velocity in the ion or electron diamagnetic direction, depending on the local values of $\eta_i$ and $Z_{\rm eff}$, is reduced to zero at the rational surface.\cite{rf3} (See Sect.~\ref{snat}.)
There are also marked differences in the predicted dependences of the shielding factors, penetration thresholds, and de-penetration thresholds, on plasma parameters between the various  models. 

It is clearly important to choose the correct response model when investigating driven magnetic reconnection in a tokamak
plasma. But, what is
the appropriate model for the problem under investigation? It is the thesis of this paper that the correct model is a
{\em two-fluid nonlinear}\/ model, and that all of the other models mentioned previously are either inadequate or invalid. 
It is obvious that a single-fluid model is inadequate, given the significant differences that exist between, for example, the 
 ${\bf E}\times
{\bf B}$ velocity and the perpendicular electron fluid velocity in a typical H-mode pedestal. 
By definition, a linear model becomes invalid as soon as the width of the magnetic separatrix exceeds the
layer width. Careful comparisons performed in Ref.~\onlinecite{uw} reveal that in typical $n=2$ DIII-D ELM suppression
experiments the driven island width at the 8/2 rational surface greatly exceeds the linear layer width in the penetrated state (which is hardly surprising). 
However, the driven island width also exceeds the linear layer width in the shielded state. This suggests that both the
penetrated and shielded states are governed by nonlinear physics. There are other strong indications of nonlinear behavior in
RMP ELM suppression experiments. For instance, Fig.~5 in Ref.~\onlinecite{naz1} shows data from an $n=2$ ELM suppression
experiment performed on DIII-D in which a bifurcation from a penetrated to a shielded state takes place. This transition is
accompanied by the ``spin-up" of the locked magnetic island chain associated with the penetrated state. Such
behavior is impossible within the context of linear response theory, but is easily explained within the context of nonlinear theory.  (See Sect.~\ref{ex3}.) Furthermore, Fig.~29 in Ref.~\onlinecite{naz2} shows a shielded
state in an $n=3$ RMP suppression experiment performed on DIII-D with a fixed RMP amplitude in an otherwise steady-state plasma. It can be seen that the state seems to consist of island
chains of pulsating width, driven at different rational surfaces in the pedestal, and rotating in highly uneven manners. Moreover, the electromagnetic torque exerted on the plasma  is time-varying, as evidenced by the time-varying ion
toroidal velocity. As before, all of these behaviors are impossible within the context of linear response theory, but are
easily accounted for within the context of nonlinear theory.\cite{rf3} (See Sect.~\ref{ex2}.)

The remainder of this paper is devoted to the exposition of a two-fluid nonlinear theory of driven magnetic reconnection at a single, tearing-stable, rational surface embedded in a
tokamak plasma. As explained in the  previous discussion, such a  theory is relevant to the problem of understanding RMP-induced ELM suppression. 

\section{Two-Fluid Nonlinear Response Model}
\subsection{Introduction}
This section describes the particular two-fluid nonlinear response model of driven magnetic reconnection at a single, tearing-stable, rational surface, embedded in a
tokamak plasma, that is adopted in this paper. 

The model in question was derived in Ref.~\onlinecite{rf3}. The core of the model is a single-helicity version of the
well-known four-field model of Hazeltine, Kotschenreuther, and Morrison.\cite{hkm} The core model is augmented by phenomenological terms
representing anomalous cross-field particle and momentum transport due to small-scale plasma turbulence. Finally, the model includes approximate (i.e.,
flux-surface averaged) expressions for the divergences of the neoclassical stress tensors. These expressions allow  neoclassical ion poloidal and perpendicular flow damping to be incorporated into the analysis. 

In Ref.~\onlinecite{rf3}, an ordering scheme is adopted that is suitable for a constant-$\psi$\,\cite{fkr}, sonic,\cite{rf3a} magnetic island chain whose radial width is similar  to  the ion
poloidal gyroradius. Momentum transport in the island region is assumed to be dominated by poloidal and perpendicular
ion neoclassical flow damping (rather than perpendicular ion viscosity). After a great deal of analysis, the formalism
reduces to a nonlinear island width evolution equation and a nonlinear island phase evolution equation. (See Sect.~\ref{model}.) These two equations are
coupled together. The formalism also determines the density, temperature, and flow profiles in the island region. (See Sects.~\ref{vely} and \ref{sprof}.)

\subsection{Magnetic Island Chain}
Consider a large aspect-ratio, low-$\beta$, circular cross-section,
 tokamak plasma equilibrium of major radius $R_0$, and toroidal magnetic field-strength $B_0$.
Let us adopt a right-handed, quasi-cylindrical, toroidal coordinate system ($r$, $\theta$, $\varphi$) whose symmetry axis ($r=0$) coincides with the
magnetic axis.  The coordinate $r$ also serves as a label
for the unperturbed (by the island chain) magnetic flux-surfaces. Let the equilibrium toroidal magnetic field and toroidal plasma current both run in the $+\varphi$ direction. 

Suppose that a helical magnetic island chain (driven by an RMP),
with $m_\theta$ poloidal periods, and $n_\varphi$ toroidal periods, is embedded in
 the aforementioned plasma. The island chain is assumed to be  radially localized in the vicinity of its
associated 
rational surface, minor radius $r_s$,  which is defined as the unperturbed magnetic flux-surface at which $q(r_s)=m_\theta/n_\varphi$. Here, $q(r)$ is the   
equilibrium safety-factor profile. 
Let the full radial width of the island chain's magnetic separatrix be $4 \,w$.
In the following, it is assumed that
$r_s/R_0\ll 1$ and $w/r_s\ll 1$. 

The magnetic flux surfaces in the island region correspond to the contours of\,\cite{rf3}
\begin{equation}
{\mit\Omega}(X,\zeta) = \frac{1}{2}\,X^{\,2}+\cos\zeta,
\end{equation}
where
$X=(r-r_s)/w$, $\zeta = \chi-\phi_p(t)$, and $\chi= m_\theta\,\theta-n_\varphi\,\varphi$.  The
O-points of the island chain are located at $X=0$ and $\zeta=\pi$, whereas the X-points
are located at $X=0$ and $\zeta=0$. The magnetic separatrix corresponds to ${\mit\Omega}=1$, the
region enclosed by the separatrix to $-1\leq {\mit\Omega}<1$, and the region outside the separatrix
to ${\mit\Omega}>1$. 

It is helpful to define the concept of a {\em vacuum island chain}, which is is defined as the static island chain
obtained by naively superimposing the vacuum resonant component of the RMP onto the unperturbed 
plasma equilibrium. Let $4\,w_v$ be the full radial width of the vacuum island chain. Moreover,
 $\phi_p$ is conveniently defined as the helical phase shift between the true and the vacuum island chains. 

\subsection{Neoclassical Flow Damping}
Let $\nu_{\theta\,i}$ and $\nu_{\perp\,i}$ be the neoclassical ion poloidal and perpendicular
flow damping rates, respectively, in the vicinity of the island chain. The relationships between these damping rates
and the assumed forms of the neoclassical stress tensors are specified in Ref.~\onlinecite{rfx}. 

Neoclassical ion poloidal flow damping acts to relax the ion poloidal flow velocity in the
vicinity of the rational surface to $-\lambda_{\theta\,i}\,\eta_i\,V_\ast$, whereas neoclassical
ion perpendicular flow damping acts to relax the ion perpendicular flow velocity in the
vicinity of the rational surface to $-\lambda_{\perp\,i}\,\eta_i\,V_\ast$. Here, $V_\ast= T_{i\,0}/(e\,B_0\,L_n)$
is the equilibrium ion diamagnetic velocity at the rational surface (due to density gradients only), $T_{i\,0}=T_i(r_s)$, $e$ is the
magnitude of the electron charge,  $L_n=-1/(d\ln n_e/dr)_{r=r_s}$, and $\eta_i =(\ln T_i/d\ln n_e)_{r=r_s}$.  Furthermore,  $T_i(r)$ is the equilibrium ion temperature profile, and $n_e(r)$  the equilibrium 
electron number density profile. The values of the dimensionless {\em neoclassical velocity parameters}, $\lambda_{\theta\,i}$ and $\lambda_{\perp\,i}$,
are specified in Appendix~\ref{appx}. 

\subsection{Natural Frequency}\label{snat}
The {\em natural frequency}, $\omega_0$, is defined as the propagation frequency (i.e., $d\phi_p/dt$) that a
naturally unstable magnetic island chain, resonant at the rational surface in question,  would have in the absence of an RMP.\cite{rf1} According to the analysis of Ref.~\onlinecite{rf3},
\begin{equation}
\omega_0 = \omega_E - [1+(1-\lambda_{\theta\,i})\,\eta_i]\,k_\theta\,V_\ast,
\end{equation}
where $\omega_E$ is the equilibrium value of the ${\bf E}\times {\bf B}$ frequency at the rational surface, and
$k_\theta = m_\theta/r_s$. Given that $1-\lambda_{\theta\,i}$ varies between $-0.173$, for a pure electron--hydrogen
plasma, and $-0.5$, for a very impure electron--hydrogen plasma (see Table~\ref{tab1}), the previous formula
implies that $\omega_0-\omega_E < 0$ unless $\eta_i$ exceeds a  critical value. This critical
value is  $5.78$ for a pure plasma,
and $2$ for a very impure plasma. (See Table~\ref{tab1}.) If $\eta_i$ does not exceed the critical value then
$\omega_0-\omega_E<0$: i.e., the natural frequency is offset from the local ${\bf E}\times {\bf B}$ frequency in the {\em ion}\/ diamagnetic
direction. On the other hand, if $\eta_i$ exceeds the critical
value then $\omega_0-\omega_E>0$: i.e., the natural frequency is   offset from the local equilibrium  ${\bf E}\times {\bf B}$ frequency in the {\em electron}\/ diamagnetic
direction.
Incidentally, the fact that magnetic island chains can propagate in the ion
diamagnetic direction relative to the local equilibrium ${\bf E}\times {\bf B}$ frame has been verified experimentally.\cite{lahaye,bur} 
Note, finally, that the physics that determines the natural frequency of a nonlinear magnetic island chain is completely different to
that which determines the rotation frequency of a linear drift-tearing mode.\cite{coppi}

The natural frequency plays a crucial role in the theory of driven magnetic reconnection at a tearing-stable rational
surface because, in order for strong shielding at the surface to break down, the local plasma flow must be modified
in such a manner that the natural frequency becomes zero. (See Sect.~\ref{ex4}.)

\subsection{Fundamental Timescales}
Let
\begin{equation}
\tau_H = \frac{2\,\sqrt{I_1\,I_v}}{(2\,m_\theta)^{\,3/2}}\left(\frac{q_s}{\epsilon_s}\right)^2 L_q\left(
\frac{\mu_0\,n_0\,m_i}{B_0^{\,2}}\right)^{1/2}
\end{equation}
be the effective {\em hydromagnetic timescale}\/ at the rational surface. Here, $q_s=m_\theta/n_\varphi$, $\epsilon_s=r_s/R_0$, $L_q=1/(d\ln q/dr)_{r=r_s}$, $n_0=n_e(r_s)$, and $m_i$  is the ion mass. The dimensionless quantities $I_1$ and $I_v$ are defined in Appendix~\ref{appu}. 

Let 
\begin{equation}
\tau_R = \frac{I_i}{6\,m_\theta}\,\frac{\mu_0\,r_s^{\,2}}{\eta_\parallel}
\end{equation}
be the effective {\em resistive diffusion timescale}\/ at the rational surface. Here, $\eta_\parallel$ is the
equilibrium parallel electrical resistivity at the rational surface. The dimensionless quantities $I_i$ is defined in Appendix~\ref{appu}. 

Finally, let 
\begin{equation}
\omega_D = \left(\frac{\epsilon_s}{q_s}\right)^{\,2}\,\frac{\nu_{\theta\,i}}{I_1} 
\end{equation}
be the effective {\em poloidal flow damping rate}\/ at the rational surface. Incidentally, according to Ref.~\onlinecite{kim}, 
\begin{equation}
\omega_D = \frac{f_t}{I_1}\,\nu_i,
\end{equation}
where $f_t$ is the fraction of trapped particles, and $\nu_i$ is the ion collision frequency. (Both quantities are evaluated at the rational surface.) Here, it is assumed that
the ions are  in the banana collisionality regime. 

\subsection{Island Evolution Equations}\label{model}
The two-fluid nonlinear response model derived in Ref.~\onlinecite{rf3} reduces to an {\em island width evolution equation},
\begin{equation}\label{e47}
\lambda_R\,\frac{d\xi}{dT} = -\xi^{\,2/3}+b_v\,\cos\phi_p,
\end{equation}
and an {\em island phase evolution equation}, 
\begin{equation}\label{e48}
\frac{d^{\,2}\phi_p}{dT^{\,2}}+\frac{d\phi_p}{dT} +b_v\,\xi^{\,1/3}\,\sin\phi_p = \gamma.
\end{equation}
Here, 
\begin{align}
T &= \omega_D\,t,\\[0.5ex]
\lambda_R&= S\,(\omega_D\,\tau_H)^{\,5/3},\label{lamr}\\[0.5ex]
S &=\frac{\tau_R}{\tau_H},\\[0.5ex]
\xi &= \hat{w}^{\,3},\\[0.5ex]
\hat{w} &=\frac{w}{w_0},\\[0.5ex]
\frac{w_0}{r_s} &= (\omega_D\,\tau_H)^{\,2/3},\label{ew0}\\[0.5ex]
b_v &= \left(\frac{w_v}{w_0}\right)^{\,2},\label{ebv}\\[0.5ex]
\gamma&=\frac{\omega_0}{\omega_D}.\label{gam}
\end{align}

In Eqs.~(\ref{e47}) and (\ref{e48}),  $\xi$ characterizes the island width (actually, it is proportional to the cube of the island width), $\phi_p$ is the helical phase of the island
chain relative to the vacuum island chain, $b_v$ is the normalized amplitude of the  resonant component of the RMP, $\lambda_R$ is the ratio of the typical
island width growth timescale to the typical island rotation timescale, and $\gamma$ is the normalized island natural frequency. 

The first term on the right-hand side of Eq.~(\ref{e47}) represents the intrinsic stability of the island
chain. (Here, it is assumed that ${\mit\Delta}'\,r_s=-2\,m_\theta$, where ${\mit\Delta}'$ is the conventional
tearing stability index.\cite{fkr}) The second term represents the  effect
of the resonant component of the RMP on island width evolution. 

The first term on the left-hand side of
Eq.~(\ref{e48}) represents ion inertia. The second term represents ion neoclassical flow damping. The
third term represents the electromagnetic locking torque due to the resonant component of the RMP.
Finally, the term on the right-hand side represents intrinsic plasma flow at the rational surface. 

Equations~(\ref{e47}) and (\ref{e48}) are highly nonlinear, and, in general, can only be solved numerically. 
Numerical integration of these equations is a relatively straightforward task, apart from one caveat. The
quantity $\xi$ cannot be negative (because the island width cannot be negative). Hence, when integrating Eqs.~(\ref{e47})
and (\ref{e48}), if $\xi$ passes through zero then the following transformation is applied:\,\cite{rf6}
\begin{align}
\xi&\rightarrow -\xi,\\[0.5ex]
\phi_p &\rightarrow \phi_p + {\rm sgn}(\gamma)\,\pi.
\end{align}
This transformation causes the island O-points to become X-points, and vice versa, which is a natural consequence of a 
reversal in sign of the reconnected magnetic flux at the rational surface.

Note that our model neglects the influence of the perturbed bootstrap current, the perturbed ion
polarization current, and magnetic field-line curvature, on island width evolution, on the assumption that
these effects are much less important than the destabilizing effect of the resonant component of the RMP. 

\subsection{Toroidal Ion Velocity Profile}\label{vely}
It is helpful to define the flux-surface label $k=[(1+{\mit\Omega})/2]^{\,1/2}$. Thus, the island
O-point corresponds to $k=0$, and the magnetic separatrix to $k=1$. Furthermore, $|X|\simeq 2\,k$ in the
limit $k\gg 1$. Let
\begin{equation}
\hat{V}_{\varphi\,i}(k) =- \frac{V_{\varphi\,i}\,(k)}{(R_0/n_\varphi)\,\omega_D},
\end{equation}
where $V_{\varphi\,i}(k)$ is the toroidal ion velocity profile. (Note that the toroidal ion velocity---or, to be more exact, the
parallel ion velocity---is a magnetic flux-surface function.\cite{rf3})
According to the analysis of Ref.~\onlinecite{rf3}, 
\begin{align}
\hat{V}_{\varphi\,i}(k<1) &= \frac{d\phi_p}{dT},\\[0.5ex]
\hat{V}_{\varphi\,i}(k>1)&= \frac{d\phi_p}{dT}+\frac{1}{(1+\bar{\nu})\,{\cal A}\,{\cal C}\,f}\left[\bar{\nu}\left(\gamma-\frac{d\phi_p}{dT}\right)
-\frac{1}{I_1}\left(1-\frac{\bar{\nu}}{f}\right)\frac{d^{\,2}\phi_p}{dT^{\,2}}\right].
\end{align}
Here, $\bar{\nu}=\nu_{\perp\,i}/\nu_{\theta\,i}$, the function $f(k)$ is defined in Appendix~\ref{appu}, and the
functions ${\cal A}(k)$ and ${\cal C}(k)$ are defined in Appendix~\ref{apa}. 

\subsection{Density and Temperature Profiles}\label{sprof}
The electron number density profile in the vicinity of the island chain is a magnetic flux-surface function that takes the form
\begin{align}
n_e(k<1) &= n_0,\\[0.5ex]
n_e(k>1)&= n_0\left[1-{\rm sgn}(X)\,\frac{w}{L_n}\int_1^k\frac{2\,dk}{\cal C}\right].
\end{align}
The ion temperature profile in the vicinity of the island chain is a magnetic flux-surface function that takes the form
\begin{align}
T_i(k<1) &= T_{i\,0},\\[0.5ex]
T_i(k>1)&= T_{i\,0}\left[1-{\rm sgn}(X)\,\frac{\eta_i\,w}{L_n}\int_1^k\frac{2\,dk}{\cal C}\right].
\end{align}
Finally, the electron temperature profile in the vicinity of the island chain is a magnetic flux-surface function that takes the form
\begin{align}
T_e(k<1) &= T_{e\,0},\\[0.5ex]
T_e(k>1)&= T_{e\,0}\left[1-{\rm sgn}(X)\,\frac{\eta_e\,w}{L_n}\int_1^k\frac{2\,dk}{\cal C}\right].
\end{align}
Here, $T_{e\,0}=T_e(r_s)$ and $\eta_e=(d\ln T_e/d\ln n_e)_{r=r_s}$, where $T_e(r)$ is the equilibrium electron temperature profile. 
Note that the density, ion temperature, and electron temperature, profiles are all flattened inside the island separatrix. 

\section{Approximate Analytic Solutions}\label{sx}
\subsection{Introduction}
Before attempting to solve Eqs.~(\ref{e47}) and (\ref{e48}) numerically, it is helpful to search for 
approximate analytic solutions of these equations.

\subsection{Renormalization}
The analytic solution of Eqs.~(\ref{e47}) and (\ref{e48}) is facilitated by defining the following rescaled variables:
\begin{align}
\hat{T}&= \gamma\,T,\\[0.5ex]
\hat{\lambda}_R&= \gamma^{\,4/3}\,\lambda_R,\label{lamr1}\\[0.5ex]
\hat{b}_v &= \frac{b_v}{\gamma^{\,2/3}},\\[0.5ex]
\hat{\xi} &= \frac{\xi}{\gamma}.
\end{align}
When re-expressed in terms of these new variables, Eqs.~(\ref{e47}) and (\ref{e48}) take the form
\begin{align}
\hat{\lambda}_R\,\frac{d\hat{\xi}}{d\hat{T}} &= -\hat{\xi}^{\,2/3}+\hat{b}_v\,\cos\phi_p,\label{e26}\\[0.5ex]
\gamma\,\frac{d^{\,2}\phi_p}{d\hat{T}^{\,2}}+\frac{d\phi_p}{d\hat{T}} +\hat{b}_v\,\hat{\xi}^{\,1/3}\,\sin\phi_p &= 1\label{e27}.
\end{align}
Note that we are assuming, without loss of generality,  that $\gamma>0$. 

\subsection{Locked Regime}\label{locked}
Let us search for a locked solution of Eqs.~(\ref{e26}) and (\ref{e27}) in which $\hat{\xi}$ and $\phi_p$ are
both constant in time (i.e., $d/d\hat{T}=0$, which implies that the island chain is stationary in the laboratory frame).  In this case, it is easily demonstrated that
\begin{align}
\hat{\xi} &= \hat{b}_v^{\,3/2}\,\cos^{3/2}\phi_p,\\[0.5ex]
\hat{b}_v^{\,3/2}\,\cos^{1/2}\phi_p\,\sin\phi_p&= 1.
\end{align}
The previous two equations reveal that locked solutions exist as long as $\hat{b}_v>\hat{b}_{v\,{\rm unlock}}$, where
\begin{equation}\label{unlock}
\hat{b}_{v\,{\rm unlock}}=\left(\frac{\sqrt{27}}{2}\right)^{1/3}= 1.374.
\end{equation}
Moreover, such solutions are characterized by $0\leq \phi_p\leq \phi_{p\,{\rm unlock}}$, where
\begin{equation}
\phi_{p\,{\rm unlock}}= \sin^{-1}\left(\sqrt{\frac{2}{3}}\right)=54.7^\circ.
\end{equation}
Finally, 
\begin{equation}
\frac{w}{w_v} = \frac{\hat{\xi}^{\,1/3}}{\hat{b}_v^{\,1/2}} = \cos^{1/2}\phi_p.
\end{equation}
Given that $\cos\phi_p > 1/\sqrt{3} = 0.5774$ for locked solutions, we deduce that the locked island
width, $w$, is similar in magnitude to the vacuum island width, $w_v$. In other words, there is  no effective shielding in the so-called {\em locked
regime}. 

\subsection{Pulsating Regime}
Let us search for a solution of Eqs.~(\ref{e26}) and (\ref{e27}) in which the island rotates in the laboratory frame (i.e., $d/d\hat{T}\neq 0$). Suppose that the term on the left-hand side of Eq.~(\ref{e26}), and the first term on the left-hand side of Eq.~(\ref{e27}), are both negligible. 
In this case, Eqs.~(\ref{e26}) and (\ref{e27}) reduce to 
\begin{align}
\hat{\xi}^{\,2/3} &= \hat{b}_v\,\cos\phi_p,\label{e33}\\[0.5ex]
\frac{d\phi_p}{d\hat{T}}+\hat{b}_v\,\hat{\xi}^{\,1/3}\,\sin\phi_p&= 1.\label{e34}
\end{align}
If the second term on the left-hand side of Eq.~(\ref{e34}) is negligible then
$d\phi_p/d\hat{T}= 1$, which justifies the neglect of the first term on the left-hand side of Eq.~(\ref{e27}).
Equation~(\ref{e33}) yields
\begin{equation}\label{e35}
\hat{\xi} = \hat{b}_v^{\,3/2}\,\cos^{3/2}\phi_p.
\end{equation}
Obviously, this solution is only valid when $\cos\phi_p \geq 0$, which implies that $-\pi/2 \leq \phi_p\leq \pi/2$. 
Moreover, 
\begin{equation}
\frac{w}{w_v} = |\cos\phi_p|^{\,1/2}.
\end{equation}
It follows that the island 
 width  pulsates,  periodically falling to zero, at which times the island helical
phase---which, otherwise, increases continually in time---jumps from $\pi/2$ to $-\pi/2$.\cite{rf2,rf6,rf3} Furthermore, because the maximum allowed value
of $\cos\phi_p$ is $1$, we deduce that there is  no effective shielding in the so-called {\em pulsating regime}\/ [i.e., $w/w_v\sim {\cal O}(1)$]. 

Let us assume that the neglect of the second term on the left-hand side of Eq.~(\ref{e34}) is justified
as long as it is not possible to find a locked solution of this equation with $\xi$ given by Eq.~(\ref{e35}). 
In other words, the neglect is justified as long as 
\begin{equation}
\hat{b}_v^{\,3/2}\,\cos^{1/2}\phi_p\,\sin\phi_p= 1.
\end{equation}
is insoluble. This is the case provided that $\hat{b}_v < \hat{b}_{v\,{\rm unlock}}$, where $\hat{b}_{v\,{\rm unlock}}$
is specified in Eq.~(\ref{unlock}). 

Finally, it is easily demonstrated that the neglect of the term on the left-hand side of Eq.~(\ref{e26}) is justified provided
\begin{equation}
\hat{b}_v< \frac{1}{\hat{\lambda}_R^{\,2}}.
\end{equation}

\subsection{Suppressed Regime}\label{suppressed}
Let us search for another rotating solution of Eqs.~(\ref{e26}) and (\ref{e27}). Suppose that the first term on the right-hand side of Eq.~(\ref{e26}), and the first term on the left-hand side of Eq.~(\ref{e27}), are both negligible. 
In this case, Eqs.~(\ref{e26}) and (\ref{e27}) reduce to 
\begin{align}
\hat{\lambda}_R\,\frac{d\hat{\xi}}{d\hat{T}}&=\hat{b}_v\,\cos\phi_p,\label{e33a}\\[0.5ex]
\frac{d\phi_p}{d\hat{T}}+\hat{b}_v\,\hat{\xi}^{\,1/3}\,\sin\phi_p&= 1.\label{e34a}
\end{align}
If the second term on the left-hand side of Eq.~(\ref{e34a}) is negligible then
$d\phi_p/d\hat{T}= 1$, which justifies the neglect of the first term on the left-hand side of Eq.~(\ref{e27}), and
Eq.~(\ref{e33a}) can be integrated to give
\begin{equation}\label{e35a}
\hat{\xi} = \frac{\hat{b}_v}{\hat{\lambda}_R}\,\sin\phi_p.
\end{equation}
Obviously, this solution is only valid when $\sin\phi_p \geq 0$ (because $\hat{\xi}$ cannot be negative), which implies that $0 \leq \phi_p\leq \pi$. 
Moreover, 
\begin{equation}\label{e42}
\frac{w}{w_v} = \frac{|\sin\phi_p|^{1/3}}{\hat{b}_v^{\,1/6}\,\hat{\lambda}_R^{\,1/3}}.
\end{equation}
It follows that the island width  pulsates, periodically falling to zero, at which times the island helical
phase---which, otherwise, increases continually in time---jumps from $\pi$ to $0$.\cite{rf2,rf6,rf3} 

Let us assume that the neglect of the second term on the left-hand side of Eq.~(\ref{e34a}) is justified
as long as it is not possible to find a locked solution of this equation with $\xi$ given by Eq.~(\ref{e35a}). 
In other words, the neglect is justified as long as 
\begin{equation}
\frac{\hat{b}_v^{\,4/3}}{\hat{\lambda}_R^{\,1/3}} \,\sin^{4/3}\phi_p=1.
\end{equation}
is insoluble. This is the case provided that $\hat{b}_v < \hat{b}_{v\,{\rm penetrate}}$, where 
\begin{equation}
\hat{b}_{v\,{\rm penetrate}}= \hat{\lambda}_R^{\,1/4}.
\end{equation}

Finally, it is easily demonstrated that the neglect of the first term on the right-hand side of Eq.~(\ref{e26}) is justified provided
\begin{equation}
\hat{b}_v>\frac{1}{\hat{\lambda}_R^{\,2}}.
\end{equation}
It follows, by comparison with Eq.~(\ref{e42}), that the so-called {\em suppressed regime}\/ is characterized by
strong shielding (i.e., $w/w_v\ll 1$). 

\subsection{Discussion}\label{dis}
\begin{figure}
\includegraphics[width=0.8\textwidth]{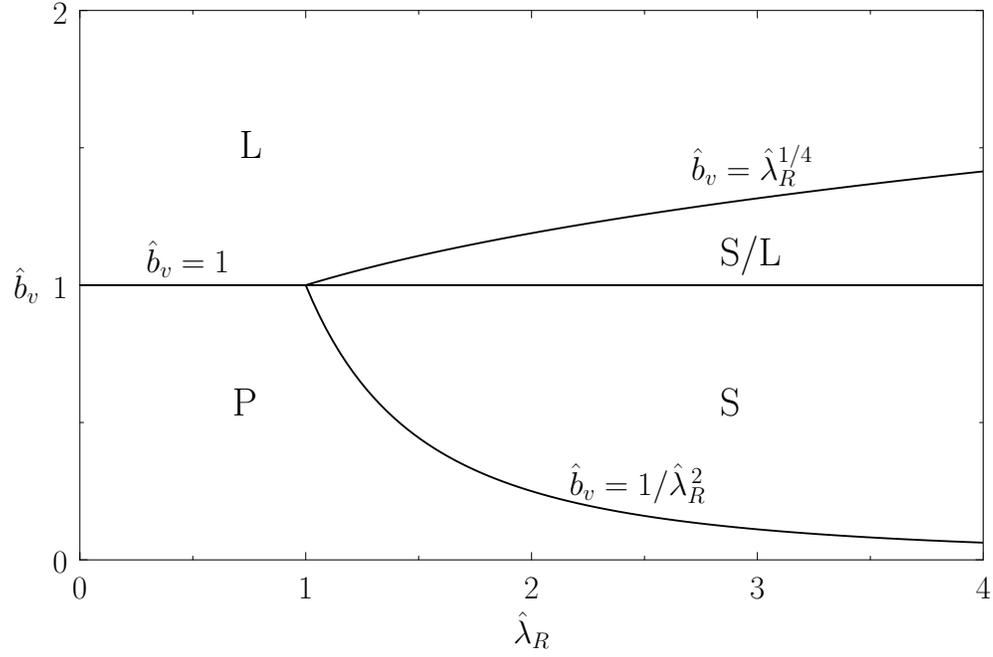}
\caption{Solution regimes for a two-fluid nonlinear response model of driven magnetic reconnection at a single, tearing-stable rational surface, embedded in a tokamak plasma, plotted in $\hat{\lambda}_R$--$\hat{b}_v$ space. L, P, and S refer to the
locked regime, the pulsating regime, and the suppressed regime, respectively.}\label{fig1}
\end{figure}

The analysis in Sects.~\ref{locked}--\ref{suppressed} lead to the scenario illustrated schematically   in Fig.~\ref{fig1}. 
As shown in the figure, there are three solution regimes in $\hat{\lambda}_R$--$\hat{b}_v$ space. Namely, the locked,
pulsating, and the suppressed regimes. Only the suppressed regime is characterized by strong shielding
(i.e., $w/w_v\ll 1$). 
Thus, referring to the discussion in Para.~8 of Sect.~\ref{s1}, the shielded state corresponds to the suppressed regime, whereas the
penetrated state corresponds to the union of the locked and pulsating regimes. 
Note that there is a region of parameter space, labelled S/L in the figure, in which
the suppressed and locked solution branches co-exist. There is a bifurcation from the suppressed to the locked
solution branch  when the upper (in $\hat{b}_v$) boundary of this region is crossed. Likewise, there is a
bifurcation from the locked to the suppressed solution branch when the lower boundary of the region is crossed. 
The former bifurcation is characterized by the sudden loss of strong shielding, whereas the latter is
characterized by the sudden onset of strong shielding. Thus, again referring to the discussion in Para.~8 of Sect.~\ref{s1}, the penetration
threshold corresponds to $\hat{b}_v>\hat{b}_{v\,{\rm penetrate}}\simeq  \hat{\lambda}_R^{\,1/4}$, whereas the de-penetration threshold
corresponds to $\hat{b}_v< \hat{b}_{v\,{\rm de-penetrate}}\simeq 1$.  

It is clear, from Fig.~\ref{fig1}, that in order to get strong shielding at the rational surface (i.e., in order to be in the
suppressed regime) it is necessary that
$\hat{\lambda}_R > 1$.
Making use of Eqs.~(\ref{lamr}), (\ref{gam}), and (\ref{lamr1}), this criterion reduces to 
$\omega_0>\omega_{0\,{\rm min}}$, where 
\begin{equation}\label{e47x}
\omega_{0\,{\rm min}}\,\tau_H = \frac{1}{S^{\,3/4}\,(\omega_D\,\tau_H)^{\,1/4}}.
\end{equation}
It follows that there is a minimum level of plasma flow at the rational surface---parameterized by the natural frequency,
$\omega_0$---required for strong shielding to be possible. 

The maximum amount of shielding in the suppressed regime is achieved at the upper
boundary of this regime in $\hat{\lambda}_R$--$\hat{b}_v$ space, which corresponds to $\hat{b}_v\simeq \hat{\lambda}_R^{\,1/4}$. It follows
from Eqs.~(\ref{lamr}), (\ref{gam}), (\ref{lamr1}), (\ref{e42}), and (\ref{e47x}) that
\begin{equation}
\left(\frac{w}{w_v}\right)_{\rm min} \simeq \frac{1}{S^{\,3/8}\,(\omega_0\,\tau_H)^{\,1/2}\,(\omega_D\,\tau_H)^{\,1/8}}=\left(\frac{\omega_{0\,{\rm min}}}{\omega_0}\right)^{1/2}.
\end{equation}

Finally,  the penetration threshold corresponds to $w_v > w_{v\,{\rm penetrate}}$, whereas the de-penetration threshold
corresponds to $w_v < w_{v\,{\rm de-penetrate}}$, where
\begin{align}
\frac{w_{v\,{\rm penetrate}}}{r_s}\simeq S^{\,1/8}\, (\omega_0\,\tau_H)^{\,1/2}\,(\omega_D\,\tau_H)^{\,3/8}= \left(\frac{\omega_0}{\omega_{0\,{\rm min}}}\right)^{1/2}\left(\frac{\omega_D\,\tau_H}{S}\right)^{1/4},\\[0.5ex]
\frac{w_{v\,{\rm de-penetrate}}}{r_s}\simeq (\omega_0\,\tau_H)^{\,1/3}\,(\omega_D\,\tau_H)^{1/3}= \left(\frac{\omega_0}{\omega_{0\,{\rm min}}}\right)^{1/3}\left(\frac{\omega_D\,\tau_H}{S}\right)^{1/4}.
\end{align}
Here, use has been made of Eqs.~(\ref{ew0}) and (\ref{ebv}).

\section{Numerical Solutions}\label{snum}
\subsection{Introduction}
Let us now consider some example numerical solutions of Eqs.~(\ref{e47}) and (\ref{e48}).

\subsection{First Example}
\begin{figure}
\includegraphics[width=0.8\textwidth]{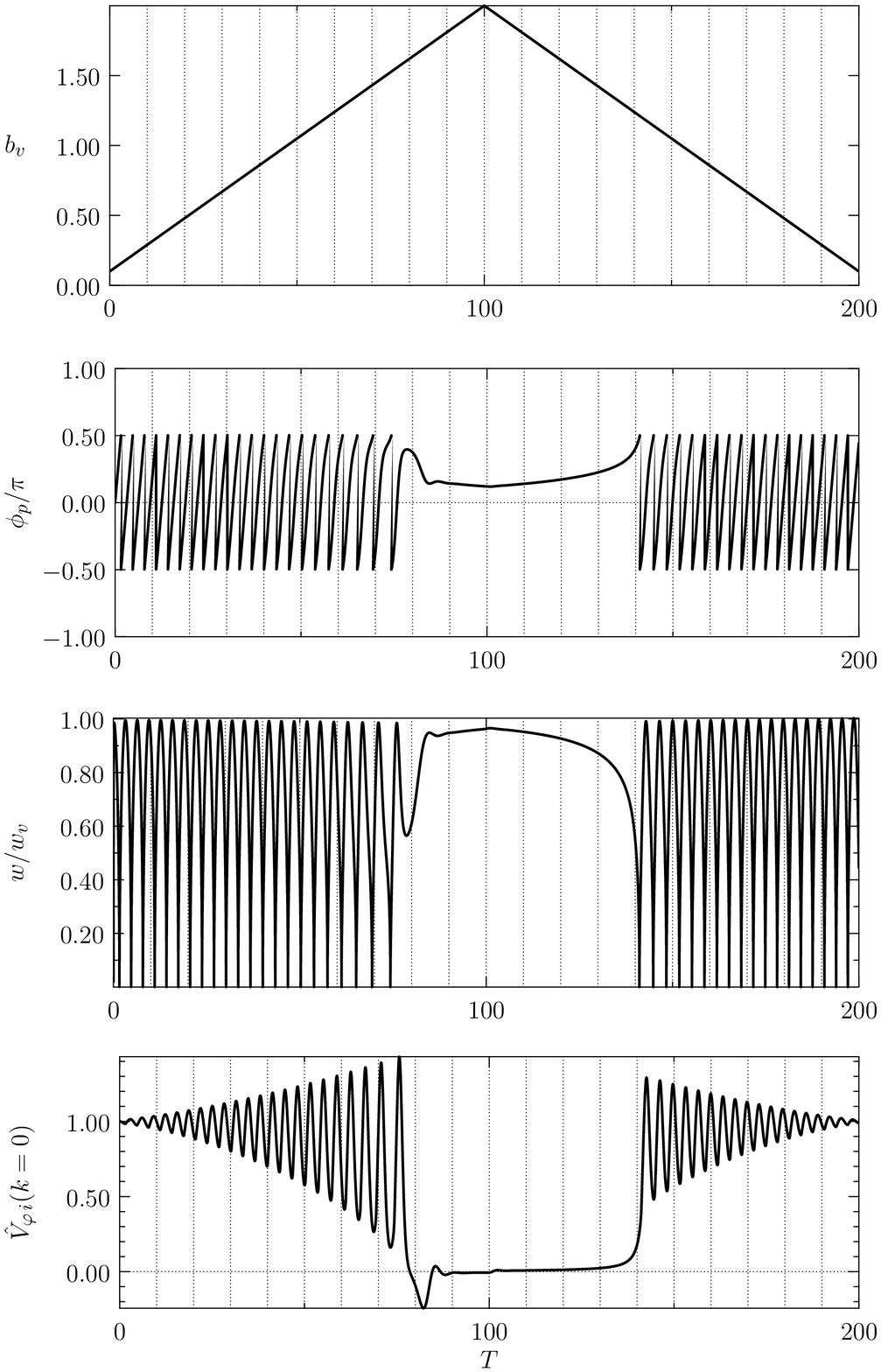}
\caption{Numerical solution of Eqs.~(\ref{e47}) and (\ref{e48}) with $\lambda_R=0.1$, $\gamma=1.0$, and $\bar{\nu}=0.1$. In order from the top
to the bottom, the panels show the normalized resonant component of the RMP, the island helical phase, the ratio of the island width to the
vacuum island width, and the normalized toroidal ion velocity inside the island separatrix. }\label{fig2}
\end{figure}

Our first example is characterized by $\lambda_R=0.1$, $\gamma=1.0$, and $\bar{\nu}=0.1$. This is a case in which the island natural frequency is not
large enough to enable strong shielding. (See Sect.~\ref{dis}.) 
The normalized resonant component
of the RMP  is increased linearly from a small value at $T=0.0$ to $b_v=2.0$ at $T=100.0$, and then decreased linearly to a small
value at $T=200.0$. Referring to Fig.~\ref{fig1},  we would expect to start off in the pulsating regime, to make a transition to the locked regime when $b_v$  exceeds  a critical value similar to unity, and then to make a back transition to the pulsating regime when $b_v$ falls below the same critical value. It can be seen, from Fig.~\ref{fig2}, that this is
essentially what happens. The pulsating regime can be identified because the helical phase of the island is restricted to the range
$-\pi/2\leq \phi_p\leq \pi/2$, the island width periodically falls to zero, and there is no shielding [i.e., $w/w_v\sim {\cal O}(1)$]. The locked regime can be identified because $\phi_p$ is relatively static, the island width has a relatively constant  nonzero value, 
$\hat{V}_{\phi\,i}(k=0)=0$ (i.e., the toroidal flow velocity inside the island separatrix is reduced to zero), and there is no shielding. Note that, in the
pulsating regime, the electromagnetic torque exerted by the resonant component of the RMP is strongly modulated, which gives rise to a modulation of
the local toroidal ion velocity. On the other hand, in the locked regime, the torque is constant, and there is no modulation of the ion velocity. Incidentally, the
modulation of the torque in the pulsating regime is an intrinsically nonlinear effect (i.e., a linear response model would give a constant torque). 
The actual transition from the pulsating to the locked regime  takes place when $b_v\simeq 1.6$, whereas the back transition takes place when
$b_v\simeq 1.2$. This is not quite what analysis presented in Sect.~\ref{sx} predicts, which is hardly surprising, given the approximate nature of the analysis. 

\begin{figure}
\includegraphics[width=0.8\textwidth]{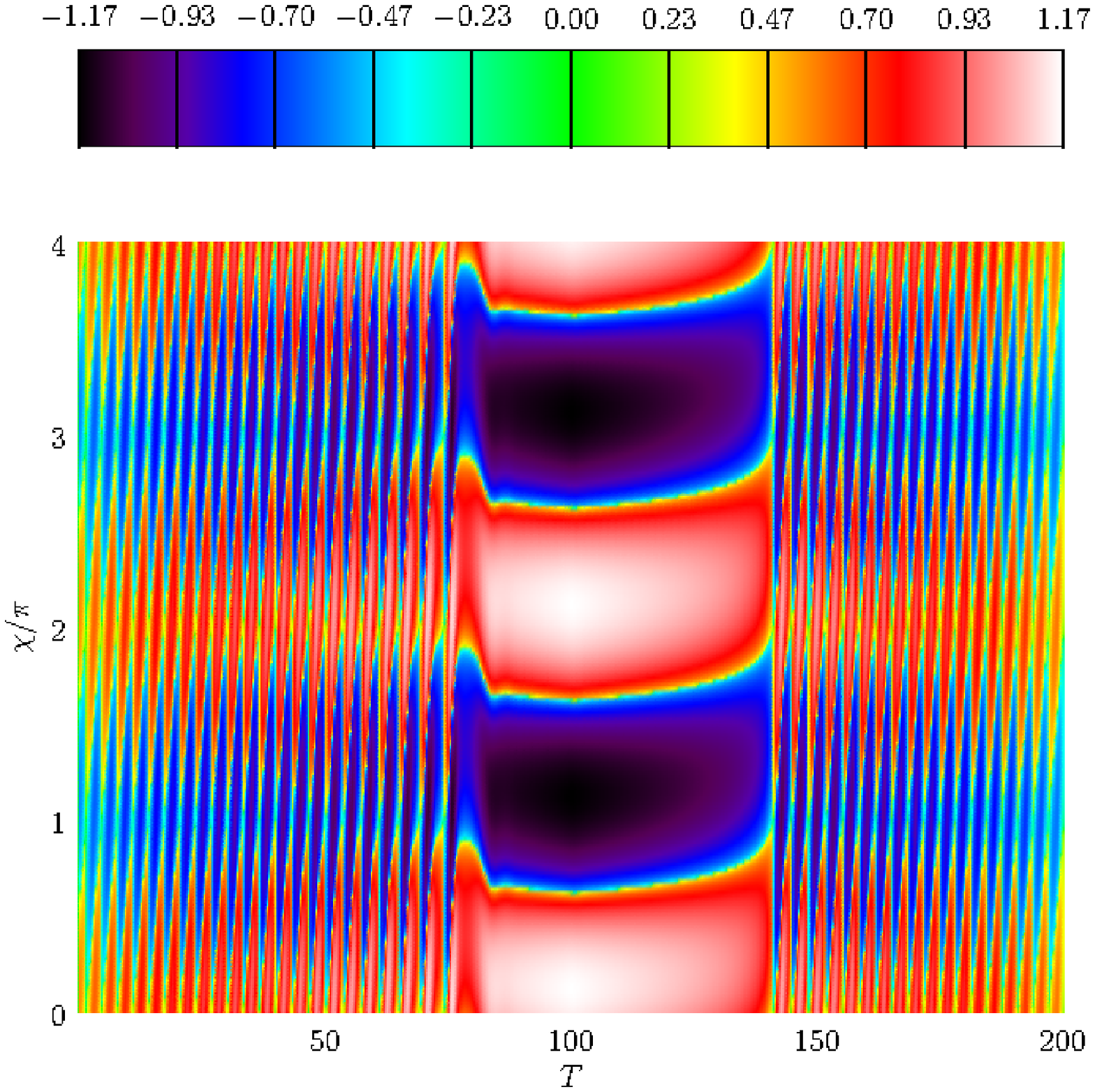}
\caption{Simulated Mirnov data for the case shown in Fig.~\ref{fig2}.}\label{fig3}
\end{figure}

Figure~\ref{fig3} shows simulated ``Mirnov" data associated with the first example. The figure actually shows contours of $b_r= w^{\,2}\,\cos(\chi-\phi_p)$ plotted in $T$-$\chi$ space (recall that $\chi=m_\theta\,\theta-n_\varphi\,\varphi$), and is meant to mimic the data that would be obtained from a comprehensive array of magnetic pick-up coils
surrounding the plasma, such as was recently installed on the DIII-D tokamak.\cite{king} The pulsating regime appears as an interlocking pattern of small regions of positive and
negative $b_r$ that are aligned almost almost parallel to the $\chi$ axis, but do not extend over all values of $\chi$.  The locked regime appears as alternating thick bands of positive and negative $b_r$ that are  aligned almost parallel to the $T$ axis. 

\subsection{Second Example}\label{ex2}
\begin{figure}
\includegraphics[width=0.8\textwidth]{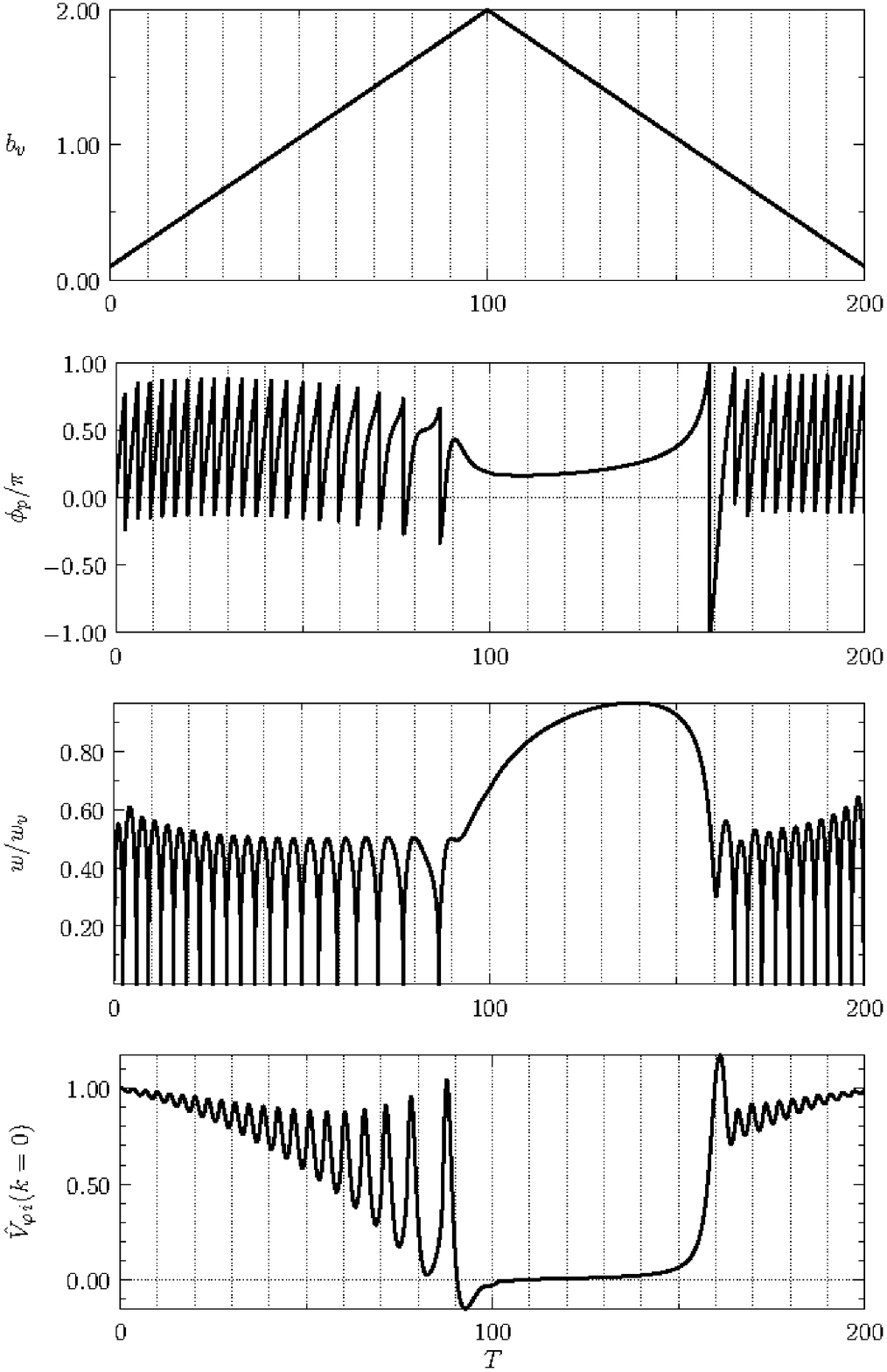}
\caption{Numerical solution of Eqs.~(\ref{e47}) and (\ref{e48}) with $\lambda_R=10.0$, $\gamma=1.0$, and $\bar{\nu}=0.1$. In order from the top
to the bottom, the panels show the normalized resonant component of the RMP, the island helical phase, the ratio of the island width to the
vacuum island width, and the normalized toroidal ion velocity inside the island separatrix.}\label{fig4}
\end{figure}

Our second example is characterized by $\lambda_R=10.0$, $\gamma=1.0$, and $\bar{\nu}=0.1$. This is a case in which the island natural frequency is
large enough to enable moderate shielding. (See Sect.~\ref{dis}.) 
The normalized resonant component
of the RMP  is increased linearly from a small value at $T=0.0$ to $b_v=2.0$ at $T=100.0$, and then decreased linearly to a small
value at $T=200.0$. Referring to Fig.~\ref{fig1},  we would expect to start off in the suppressed regime, to make a transition to the locked regime when $b_v$  exceeds a critical value somewhat larger than unity, and then to make a back transition to the suppressed  regime when $b_v$ falls below a second critical value that is similar to
unity. It can be seen, from Fig.~\ref{fig4}, that this is
essentially what happens. The suppressed regime can be identified because the helical phase of the island is restricted to the range
$0\leq \phi_p\leq \pi$, the island width periodically falls to zero, and there is moderate shielding (i.e., $w/w_v\leq 0.6$). As before, the locked regime can be identified because $\phi_p$ is relatively static, the island width has a relatively constant  nonzero value, 
$\hat{V}_{\phi\,i}(k=0)=0$, and there is no shielding.  Note that, in the
suppressed regime, the electromagnetic torque exerted by the resonant component of the RMP is strongly modulated, which gives rise to a modulation of
the local toroidal ion velocity.  As before, the
modulation of the torque in the suppressed regime is an intrinsically nonlinear effect (i.e., a linear response model would give a constant torque). 
 Note, finally, that the driven island chain makes a full rotation during the back transition from the
locked to the suppressed regimes; this is a vestigial version of the spin up described Sect.~\ref{ex3}. 

\begin{figure}
\includegraphics[width=0.8\textwidth]{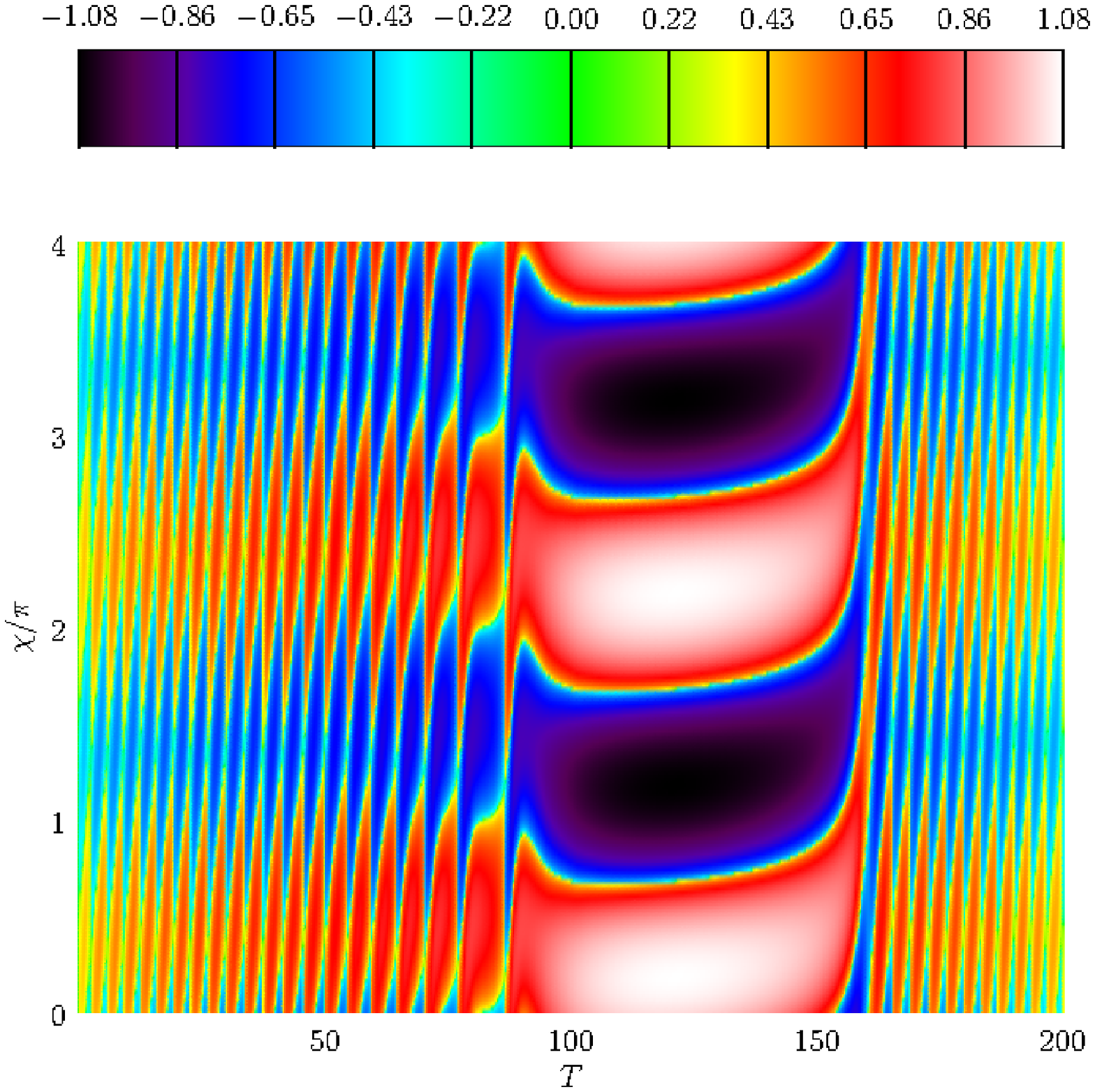}
\caption{Simulated Mirnov data for the case shown in Fig.~\ref{fig4}.}\label{fig5}
\end{figure}

Figure~\ref{fig5} shows simulated Mirnov data associated with the second example. The suppressed regime appears as an interlocking pattern of small regions of positive and
negative $b_r$ that are aligned almost parallel to the $\chi$ axis, but do not extend over all values of $\chi$.  As before, the locked regime appears as alternating thick bands of positive and negative $b_r$ that are aligned almost parallel to the $T$ axis.  

\subsection{Third Example}\label{ex3}
\begin{figure}
\includegraphics[width=0.8\textwidth]{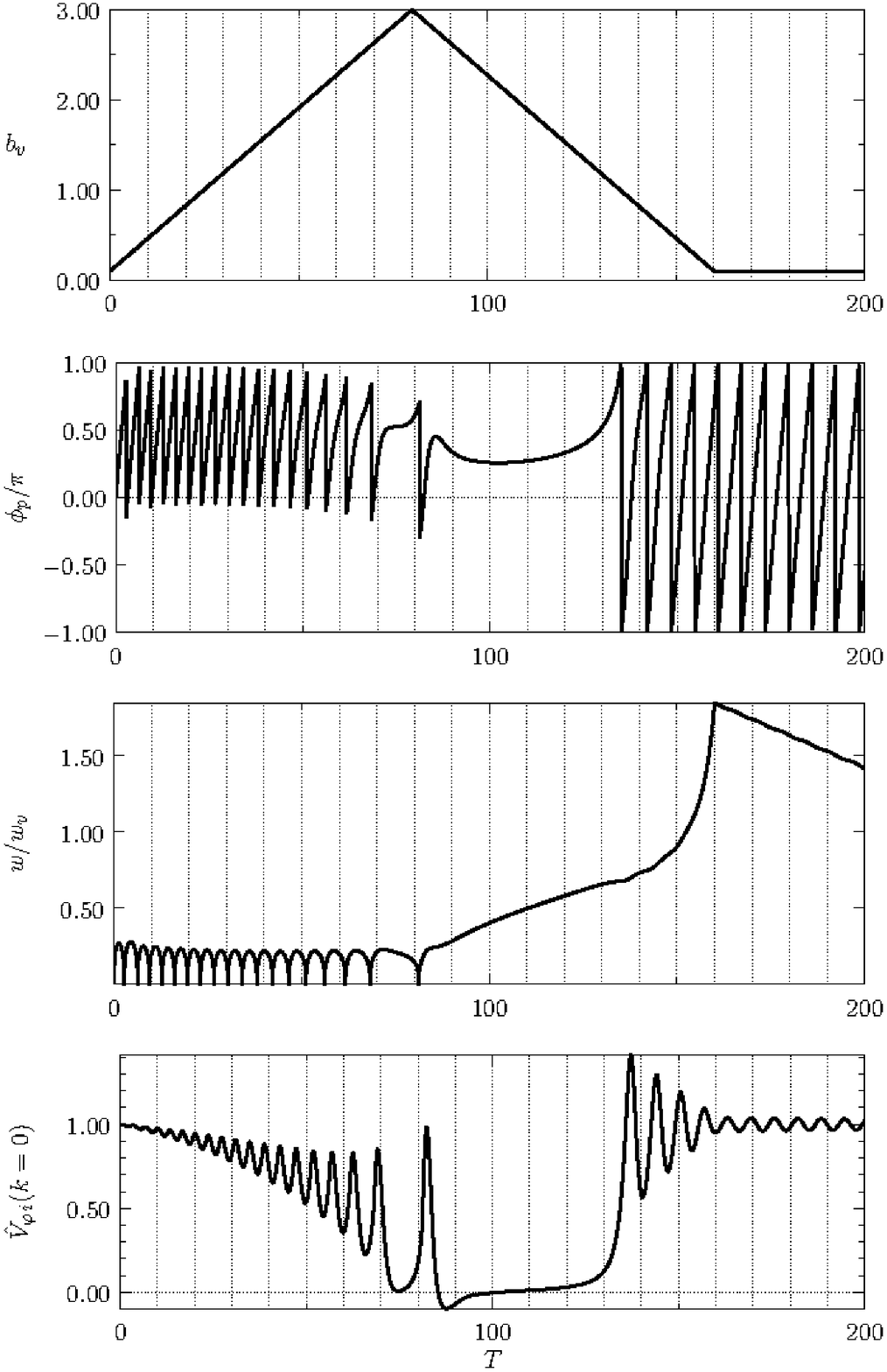}
\caption{Numerical solution of Eqs.~(\ref{e47}) and (\ref{e48}) with $\lambda_R=100.0$, $\gamma=1.0$, and $\bar{\nu}=0.1$. In order from the top
to the bottom, the panels show the normalized resonant component of the RMP, the island helical phase, the ratio of the island width to the
vacuum island width, and the normalized toroidal ion velocity inside the island separatrix.}\label{fig6}
\end{figure}

Our third example is characterized by $\lambda_R=100.0$, $\gamma=1.0$, and $\bar{\nu}=0.1$. This is a case in which the island natural frequency is
large enough to enable strong shielding. (See Sect.~\ref{dis}.)  The normalized resonant component
of the RMP  is increased linearly from a small value at $T=0.0$ to $b_v=3.0$ at $T=80.0$, decreased linearly to a small
value at $T=160.0$, and, thereafter, held steady. Referring to Fig.~\ref{fig1},  we would expect to start off in the suppressed regime, to make a transition to the locked regime when $b_v$  exceeds a critical value that is considerably larger than unity, and then to make a back transition to the suppressed  regime when $b_v$ falls below a second critical value that is similar to unity.  It can be seen, from Fig.~\ref{fig6}, that this is
essentially what happens, with one caveat (involving the spin up).  The suppressed regime can be identified because the helical phase of the island is restricted to the range
$0\leq \phi_p\leq \pi$, the island width periodically falls to zero, and there is strong shielding (i.e., $w/w_v\leq 0.25$). As before, the locked regime can be identified because $\phi_p$ is relatively static, the island width has a relatively constant  nonzero value, 
$\hat{V}_{\phi\,i}(k=0)=0$, and there is no shielding. Note, however, that at the end of the locked phase, instead of immediately re-entering the suppressed regime, the
island chain spins up: i.e., its helical phase increases continually in time. This behavior occurs because the island chain cannot decay away fast enough to prevent it from being entrained by the re-accelerated plasma
flow at the rational surface. Of course, the island chain will eventually re-enter the suppressed regime, but only when enough time has elapsed for its width to decay to
zero: i.e., after 100, or so, normalized time units.

\begin{figure}
\includegraphics[width=0.8\textwidth]{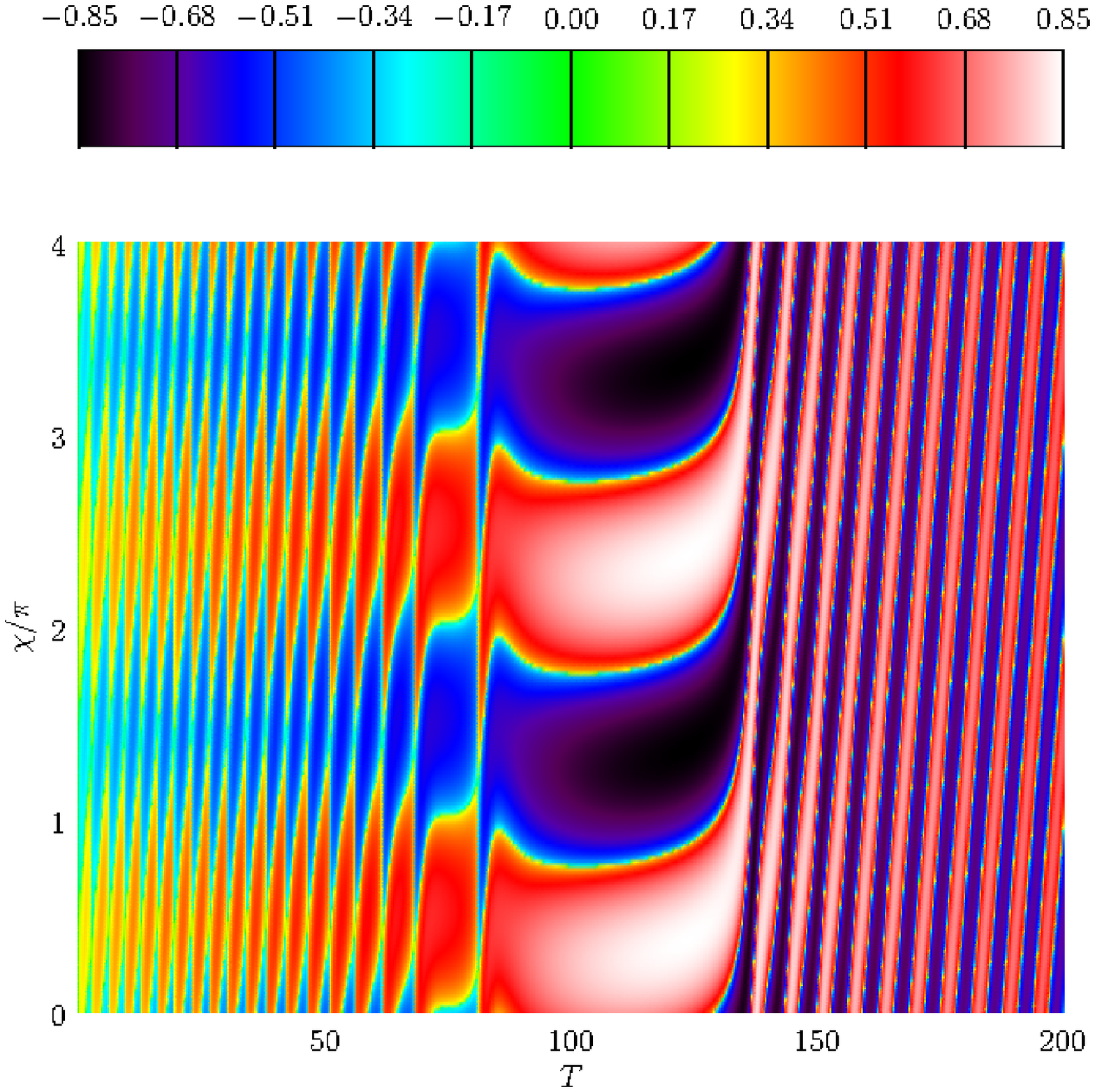}
\caption{Simulated Mirnov data for the case shown in Fig.~\ref{fig6}.}\label{fig7}
\end{figure}

Figure~\ref{fig7} shows simulated Mirnov data associated with the third example. As before, the suppressed regime appears as an interlocking pattern of small regions of positive and
negative $b_r$ that are aligned almost parallel to the $\chi$ axis, but do not extend over all values of $\chi$. However, in this particular example, it is much easier to see  that the positive and negative regions of the suppressed regime predominately occupy the same ranges
of $\chi$ as the positive and negative bands in the locked regime. 
 As before, the locked regime appears as alternating thick bands of positive and negative $b_r$ that are aligned almost parallel to the $T$ axis. Finally, the spin up
 appears as alternating diagonal bands of positive and negative $b_r$ that extend over all values of $\chi$. 
 
\subsection{Fourth Example}\label{ex4}
\begin{figure}
\includegraphics[width=0.8\textwidth]{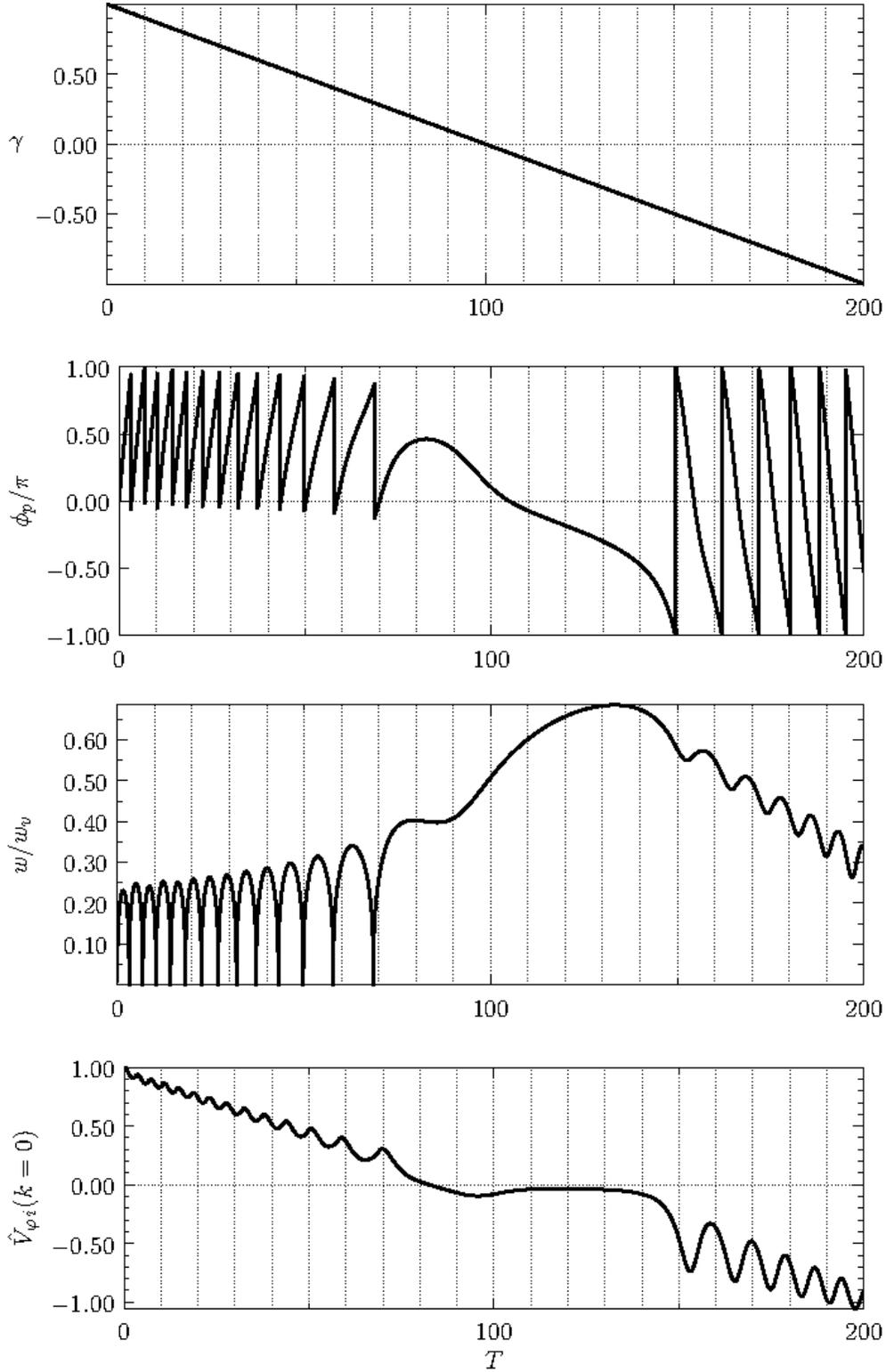}
\caption{Numerical solution of Eqs.~(\ref{e47}) and (\ref{e48}) with $\lambda_R=100.0$, $b_v=0.6$, and $\bar{\nu}=0.1$. In order from the top
to the bottom, the panels show the normalized natural frequency, the island helical phase, the ratio of the island width to the
vacuum island width, and the normalized toroidal ion velocity inside the island separatrix.}\label{fig8}
\end{figure}

\begin{figure}
\includegraphics[width=0.8\textwidth]{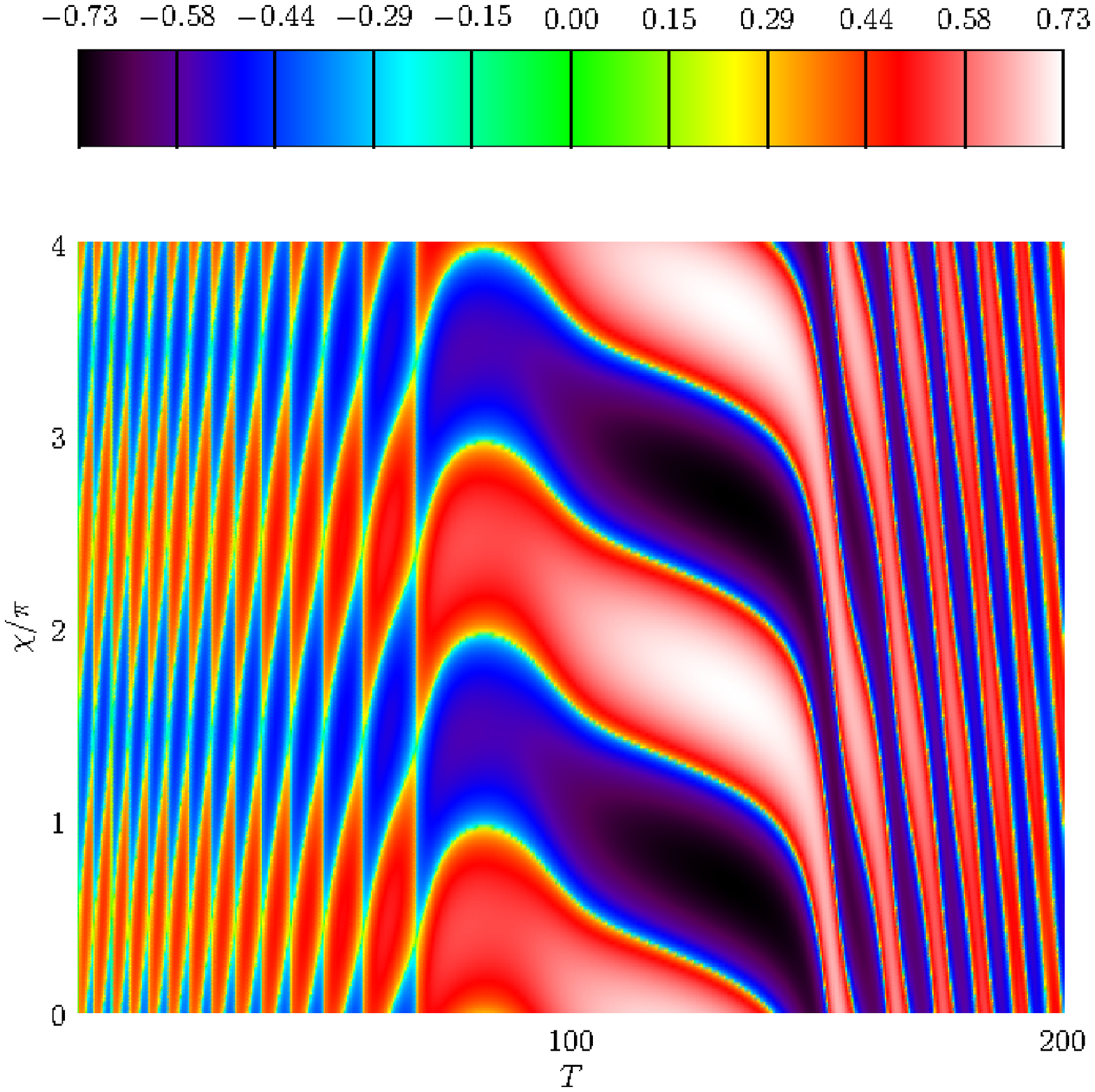}
\caption{Simulated Mirnov data for the case shown in Fig.~\ref{fig8}.}\label{fig9}
\end{figure}

Our fourth example is characterized by $\lambda_R=100.0$, $b_v=0.6$, and $\bar{\nu}=0.1$. This is a case in which the island natural frequency is
initially large enough to enable strong shielding. (See Sect.~\ref{dis}.)  The normalized island natural frequency
is ramped  linearly from 
$\gamma=1.0$ at $T=0.0$ to $\gamma=-1.0$ at $T=200.0$. This particular example is designed to illustrate what happens
when the natural frequency at the rational surface passes through zero. As can be seen from Figs.~\ref{fig8} and \ref{fig9}, we start off in the suppressed regime, there is a transition
to the locked regime when $\gamma$ becomes sufficiently small, and then when $|\gamma|$ becomes sufficiently large the locked island chain spins up
and decays aways. The only major difference between this example and the previous one is that the island spins up to a negative rotation frequency, because $\gamma$
has become negative by the time the locked island chain unlocks. Note, finally, that the locked phase is centered on the time
at which the normalized natural frequency, $\gamma$, passes through zero. In other words, the breakdown of strong shielding is  clearly 
associated with island natural frequency passing through zero. 

\section{Summary and Discussion}
The aim of this paper is to develop a physical understanding of the dynamics of magnetic reconnection
driven at a single, tearing-stable,  rational surface that is resonant with one of the dominant helical harmonics of an applied RMP. 
 The rational surface is assumed to be located just outside, but close to the top of, the pedestal of an H-mode tokamak plasma.
 Over the years, many different theoretical approaches have been taken to investigate driven reconnection at
 an intrinsically stable rational surface embedded in a tokamak plasma. These include single-fluid linear response models, two-fluid linear response models, kinetic linear response models, single-fluid nonlinear response models, and two-fluid nonlinear response models. 
  However, it is the thesis of this paper that the correct response model is a
two-fluid nonlinear model, and that all of the other models mentioned previously are either inadequate or invalid. (See the discussion in Sect.~\ref{s1}.) 

The two-fluid nonlinear response model discussed in this paper consists of an  island width evolution equation and an island phase evolution equation. (See Sect.~\ref{model}.) These two equations are
coupled together. The island width and  phase evolution equations are sufficiently nonlinear that they can only be solved accurately by numerical means. 
However, it is possible to find approximate analytical solutions of these equations. (See Sect.~\ref{sx}.) These analytic solutions reveal that there
are three different response regimes---namely, the locked, pulsating, and suppressed regimes. In the locked regime, the magnetic island chain driven at the
rational surface has a constant phase relative to the resonant component of the RMP: i.e., the island chain is stationary in the laboratory frame. Moreover, the
width of the island chain is similar to the vacuum island width, which implies that there is no effective ``shielding" (i.e., suppression of
driven magnetic reconnection) in this regime. In the pulsating and suppressed regimes, the driven island chain is forced to rotate by plasma
flow at the rational surface. However, the island width periodically falls to zero in both regimes, at which times the helical phase of the
island chain jumps by $\pi$ radians. In both regimes, the electromagnetic torque exerted by the resonant component of the
RMP is strongly modulated, which gives rise to a modulation of the local toroidal ion velocity. The main difference between the pulsating and the
suppressed regimes is that there is no effective shielding in the former regime (i.e., the driven island width is similar to the vacuum island
width), whereas there is strong shielding in the latter regime (i.e., the driven island width is much smaller than the vacuum island
width). There exists a region of parameter space in which the suppressed and the locked solution branches co-exist. Bifurcations from
one solution branch to the other are triggered when the  boundaries of this region are crossed. These bifurcations are  characterized by the sudden loss of strong shielding,  or the  sudden onset of strong shielding. 

Numerical integration of the island width and island phase evolution equations  yields results that are consistent with the aforementioned approximate analytic solutions, with one
proviso. (See Sect.~\ref{snum}.) Namely, that the transition from the locked regime to the  suppressed regime is characterized by an intermediate
regime in which the 
island chain spins up: i.e., its helical phase increases continually in time. This behavior occurs because the island chain cannot decay away fast enough to prevent it from being entrained by the re-accelerated plasma
flow at the rational surface. 

\section*{Acknowledgements}
This research was funded by the U.S.\ Department of Energy under contract DE-FG02-04ER-54742.

\appendix

\section{Neoclassical Velocity Coefficients}\label{appx}
According to Ref.~\onlinecite{kim}, we can write
\begin{equation}
\lambda_{\theta\,i}  = \frac{5}{2}-\frac{1/\sqrt{2}+\alpha}{\sqrt{2}-\ln(1+\sqrt{2})+\alpha},
\end{equation}
where $\alpha=Z_{\rm eff}- 1$. Here, $Z_{\rm eff}$ is the conventional measure of plasma impurity content.\cite{wesson} 
Furthermore, it is assumed that the majority ions have charge number unity, and are in the banana collisionality regime. 

According to Ref.~\onlinecite{shiang}, we can write
\begin{equation}
\lambda_{\perp i} = \left.\int_0^\infty \frac{x^{\,9}\,{\rm e}^{-x^{\,2}}\,(x^{\,2}-5/2)\,dx}{F(x)+\alpha}\right/
\int_0^\infty \frac{x^{\,9}\,{\rm e}^{-x^{\,2}}\,dx}{F(x)+\alpha},
\end{equation}
where $F(x)={\mit\Phi}(x)-G(x)$, $G(x)=[{\mit\Phi}(x)-x\,{\mit\Phi}'(x)]/(2\,x^{\,2})$, ${\mit\Phi}(x)$ is a standard
error function,  and $'$ denotes a derivative with respect to argument. Here, it is assumed that the majority ions have charge number unity, and are in the
$1/\nu$ collisionality regime. 

Table~\ref{tab1} illustrates the dependence of the neoclassical
velocity parameters, $\lambda_{\theta\,i}$ and $\lambda_{\perp\,i}$, on  $Z_{\rm eff}$. 

\begin{table}
\centering
\begin{tabular}{llll}\\[0.5ex]\hline
$Z_{\rm eff}$~~& $\lambda_{\theta\,i}$~~~~ & $\lambda_{\perp\,i}$~~~~~&$\eta_{i\,{\rm crit}}$\\[0.5ex]\hline
$1.0$ &$1.173$&$2.367$&$5.78$\\[0.5ex]
$2.0$ &$1.386$&$2.440$&$2.59$\\[0.5ex]
$3.0$ &$1.431$&$2.461$&$2.32$ \\[0.5ex]
$4.0$ &$1.451$& $2.471$&$2.22$\\[0.5ex]
$5.0$&$1.462$&$2.477$&$2.17$\\[0.5ex]
$\infty$&$1.5$&$2.5$&$2.0$\\[0.5ex]\hline
\end{tabular}
\caption{Neoclassical velocity coefficients as functions of $Z_{\rm eff}$. $\eta_{i\,{\rm crit}}=1/(\eta_{\theta\,i}-1)$ is the critical
value of $\eta_i$ at the rational surface above which the natural frequency is offset from the local ${\bf E}\times{\bf B}$ frequency in the
electron diamagnetic direction, as opposed to the ion diamagnetic direction.}\label{tab1}
\end{table}

\section{Useful Integrals}\label{appu}
Let
\begin{align}
I_i &= \int_0^\infty \frac{64\,[(k^{\,2}-1/2)\,{\cal A}-k^{\,2}\,{\cal C}]^{\,2}}{{\cal A}}\,dk,\\[0.5ex]
I_v &= \left(\frac{\bar{\nu}}{1+\bar{\nu}}\right) \int_1^\infty \frac{8\,({\cal A}\,{\cal C}-1)}{{\cal A}\,{\cal C}^{\,2}\,f}\,dk,\\[0.5ex]
I_1&= \frac{1}{I_v}\int_1^\infty \frac{8\,({\cal A}\,{\cal C}-1)^{\,2}}{{\cal A}\,{\cal C}^{\,2}\,f^{\,2}}\,dk,
\end{align}
where $\bar{\nu}= \nu_{\perp\,i}/\nu_{\theta\,i}$,  and
\begin{equation}
f(k) = 1-\frac{1}{1+\bar{\nu}}\,\frac{1}{{\cal A}\,{\cal C}}.
\end{equation}
Here, the functions ${\cal A}(k)$ and ${\cal C}(k)$ are defined in Appendix~\ref{apa}. 

It is easily demonstrated that\,\cite{rf3}
\begin{align}
I_i &= 3.2908,\\[0.5ex]
I_v&=\left\{\begin{array}{lll} 2^{\,1/4}\,\pi\,\bar{\nu}^{\,3/4}&\mbox{\hspace{1cm}}&\bar{\nu}\ll1\\[0.5ex]
0.35724&&\bar{\nu}\gg 1\end{array}\right.,\\[0.5ex]
I_1&=\left\{\begin{array}{lll} 0.75/\bar{\nu}&\mbox{\hspace{1.45cm}}&\bar{\nu}\ll1\\[0.5ex]
0.18182&&\bar{\nu}\gg 1\end{array}\right..
\end{align}

\section{Useful Functions}\label{apa}
\begin{align}
{\cal A}(k<1) &=\left(\frac{2}{\pi}\right)k\,K\!\left(k\right),\\[0.5ex]
{\cal A}(k>1) &=\left(\frac{2}{\pi}\right)K\!\left(\frac{1}{k}\right),\\[0.5ex]
{\cal C}(k<1) & =\left(\frac{2}{\pi}\right)\frac{\left[E\!\left(k\right)+(k^{\,2}-1)\,K(k)\right]}{k},\\[0.5ex]
{\cal C}(k>1) &=\left(\frac{2}{\pi}\right)E\!\left(\frac{1}{k}\right).
\end{align}
Here,
\begin{align}
E(x) &= \int_0^{\pi/2}(1-x^{\,2}\,\sin^2 u)^{1/2}\,du,\\[0.5ex]
K(x) &= \int_0^{\pi/2}(1-x^{\,2}\,\sin^2 u)^{-1/2}\,du
\end{align}
are standard  complete elliptic integrals.

\newpage

\end{document}